\documentstyle[prb,aps,multicol,psfig]{revtex}
\begin{document}
\draft

\title{Density Profiles, Casimir Amplitudes and Critical Exponents in the
Two Dimensional Potts Model: A Density Matrix Renormalization Study}
\author{Enrico Carlon}
\address{Institute for Theoretical Physics, Katholieke Universiteit Leuven,
Celestijnenlaan 200D, B-3001 Leuven, Belgium}
\author{Ferenc Igl\'oi}
\address{Research Institute for Solid State Physics, 
H-1525 Budapest, P.O.Box 49, Hungary\\
Institute for Theoretical Physics, Szeged University, H-6720 Szeged, Hungary}

\date{Preprint KUL-TF-97/25}

\maketitle

\begin{abstract}
We use the density matrix renormalization group (DMRG) to perform a detailed study 
of the critical properties of the two dimensional $Q$ state Potts model, including 
the magnetization and energy-density profiles, bulk and surface critical exponents 
and the Casimir amplitudes. We apply symmetry breaking boundary conditions to a $L 
\times \infty$ strip and diagonalize the corresponding transfer matrix for a series 
of moderately large systems ($L \le 64$) by the DMRG method.
The numerically very accurate finite lattice results are then extrapolated by efficient 
sequence extrapolation techniques. The critical density profiles and the Casimir 
amplitudes are found to follow precisely the conformal predictions for $Q=2$ and $3$. 
Similarly, the bulk and surface critical exponents of the models are in very good 
agreement with the conformal and exact values: their accuracy has reached or even exceeded 
the accuracy of other available numerical methods. For the $Q=4$ model both the profiles 
and the critical exponents show strong logarithmic corrections, which are also studied.
\end{abstract}


\newcommand{\bc}{\begin{center}}
\newcommand{\ec}{\end{center}}
\newcommand{\be}{\begin{equation}}
\newcommand{\ee}{\end{equation}}
\newcommand{\beqn}{\begin{eqnarray}}
\newcommand{\eeqn}{\end{eqnarray}}

\begin{multicols}{2}\narrowtext
\section{Introduction}
\label{sec:int}

The density matrix renormalization group (DMRG) method by White \cite{whitePRL} 
has considerably enhanced our abilities to diagonalize numerically the Hamiltonians
of one-dimensional quantum systems. Among the systems treated so far by the DMRG 
method we mention Heisenberg spin chains \cite{heischains}, Heisenberg ladders 
\cite{heisladders} and strongly correlated electron systems \cite{stronglyc}.
Recently, the first steps have been made towards the generalization of the method 
to other directions, such as two dimensional (2d) quantum systems \cite{2dquantum}, 
non-Hermitian Hamiltonians \cite{peschel}, etc.

According to numerical observations the DMRG method has a fast convergence,
if the ground state and the excited states of the Hamiltonian are well
separated, especially when the ground state wave function can be approximated
as a product of tensors \cite{ostlund}. This happens, among others, at the 
valence bond solid point of the generalized antiferromagnetic $S=1$ Heisenberg 
chain \cite{ostlund,VBS}. 
On the other hand the convergence of the method becomes slower at the critical 
region. The origin of this slowing down is two-fold. First, the spectrum of 
the density matrix at the critical point becomes more dense and therefore more 
states have to be kept to guarantee a given accuracy. Second, the ground state of 
the Hamiltonian is nearly degenerate in the critical region and therefore the 
computational time needed to perform the L\'anczos diagonalization increases 
considerably.

In the first application of the DMRG to the critical region \cite{andrzejDMRG} 
the couplings of the Hamiltonian were subject of a renormalization procedure 
and the critical exponents, obtained from the fixed-point transformation, 
were less accurate than those obtained by the standard renormalization group.
Later studies, concentrated on two dimensional classical systems 
\cite{nishino1,ourPRL,19V}, have demonstrated that the thermodynamic properties 
of the critical systems can be accurately obtained 
by the DMRG method. In a 2d classical system it is natural to consider the 
row-to-row transfer matrix and calculate its leading eigenvalues by the DMRG 
method, as done first by Nishino \cite{nishino1}.
More recently, the DMRG method and Baxter's corner transfer matrix
\cite{baxter} have been combined in an efficient iterative algorithm 
\cite{cornerTM}.

In this article we use the DMRG method to study the critical properties of
the two dimensional $Q$ state Potts model, in particular we focus on the 
behavior of local densities, as the magnetization and the energy density,
and on the Casimir amplitudes for $L \times \infty$ strips with symmetry
breaking boundary conditions. These quantities have been derived from 
conformal invariance, but to our knowledge, never computed directly 
from a lattice calculation.
We also calculate the complete set of surface and bulk critical exponents 
and compare our results with the exact ($Q=2$) and conjectured values 
($Q=3,4$) for these exponents.

In the usual DMRG procedure the thermodynamic limit is approached by simply 
increasing the size of the system; typically one considers quantum chains of 
lengths of the order of $10^3$, which are generated by many DMRG iterations. 
Quantities as order parameters and correlation lengths are calculated from 
such large lattices. 
The drawback of this approach is that for such large systems the accuracy of 
the DMRG method is not as good as for chains of length of one order of magnitude 
smaller, especially at the critical point, for the reasons discussed above.
On the other hand in the traditional finite size scaling analysis \cite{barber}, 
which is based on a numerically exact diagonalization of the Hamiltonian by the 
L\'anczos method, the size of the finite lattices are of the order of $10$, 
which is often too small value for an accurate determination of critical 
exponents from finite size scaling.

In the present study we combine the standard finite size scaling analysis with 
the computational power of the DMRG method. Using White's finite system
algorithm \cite{whitePRB} we calculate quantities with very high accuracy 
for strips of moderate widths ($L \leq 64$). The advantage with respect to the 
exact diagonalization of small lattices is that the finite size scaling is 
performed on a wider range of system sizes, corrections to scaling have weaker 
effects and one can determine critical exponents with a high accuracy. 
Through the finite system DMRG algorithm one is also able to calculate
accurately not only bulk, but at the same time also surface exponents, 
obtaining thus a deeper knowledge of the critical properties of the system.

The structure of the paper is the following. In Section \ref{sec:mod} we 
describe the DMRG method and its application to the $Q$ state Potts model. 
Our results about the critical density profiles are presented in Section 
\ref{sec:pro}, whereas the critical exponents and the Casimir amplitudes 
are investigated in Section \ref{sec:crit}. Finally, in Section \ref{sec:disc} 
we discuss our results.

\section{The DMRG method and its application to the Potts model}
\label{sec:mod}

The $Q$ state Potts model\cite{wu} - due to many exact and conjectured results 
- is an important testing ground for different approximate theories and numerical
methods in statistical physics. 
We consider a $L \times \infty$ lattice labeled by $l = 1$, $2$ 
\ldots $L$ and $-\infty < k < \infty$ with an Hamiltonian written as

\beqn
\beta H=-K \sum_{k} \sum_{l=1}^L \left\{
\delta \left(s_{l,k} - s_{l+1,k} \right) -
\delta \left(s_{l,k} - s_{l,k+1} \right)
\right\}
\label{ham}
\eeqn
in terms of a $Q$ component spin variable $s_{l,k} =0$, $1$, \dots,
$Q-1$ ($\beta$ denotes the inverse temperature). 
In the following we take $\beta=1$ and consider the 
ferromagnetic model ($K>0$).
In this case one can obtain the phase transition 
point of the system 
\beqn
e^{K} = 1 + \sqrt{Q},
\label{critp}
\eeqn
and the energy per bond at the transition point
\begin{eqnarray}
\epsilon_0 = \frac{1}{2} \left(1 +\frac{1}{\sqrt Q} \right).
\end{eqnarray}
from a duality transformation\cite{wu}. For general $Q$ the model is
integrable {\it at the transition point} and the transition is of
second (first) order for $Q \le 4$ ($Q>4$)\cite{baxterpotts}. In this paper we restrict
ourselves to the second order transition regime.

Since the model - except of the case of the $Q=2$ (Ising model) - is not
solved outside of the critical point the critical exponents are not
known exactly for general $Q$. Their values, however, are conjectured by
conformal field theory \cite{dotsenko,cardy,cardy++} and by approximate
mappings \cite{conjectured}.

In a $L \times \infty$ strip the local densities, such as the magnetization
\begin{eqnarray}
m_L (l) = \frac{Q \langle \delta(s_{l,k}) \rangle -1}{Q-1},
\label{magnet}
\end{eqnarray}
and the singular part of the energy density
\begin{eqnarray}
\epsilon_L(l) = \langle \delta(s_{l,k} - s_{l+1,k})\rangle - \epsilon_0,
\label{energy}
\end{eqnarray}
depend on the position of the layers $l=1$, $2$ \dots $L$. 

In the transfer matrix formalism the elements of the row-to-row transfer matrix 
$T_L$ are labeled by the possible states of a row in the system:
$i = 1$, $2$, \ldots $Q^L$.
The leading eigenvalue $\lambda_L$ of the transfer matrix
\beqn
T_L \, | v_L \rangle &=& \lambda_L \, | v_L \rangle,
\label{eigex}
\eeqn
gives the free energy per spin as
\beqn
f_L = - {1 \over L} \ln \lambda_L.
\eeqn
If the dominant eigenvector $|v_L\rangle$ is normalized, the squares of
its elements $v_L^2(i)$ are equal to the probabilities of finding a row 
in a state $i$.
Average quantities as $\langle \delta(s_{l,k}) \rangle$ can be calculated
then:
\be
\langle \delta(s_{l,k}) 
\rangle = \sum_{i=1}^{Q^L} \, v_L^2 (i) \,\, \delta(s_l(i)),
\ee
where $s_l(i)$ denotes the value of the $l$-th spin for a row in the $i$-th 
configuration.
In this way one derives magnetization and energy density profiles from the
dominant eigenvector $| v_L \rangle$.

The problem of the transfer matrix approach is that a numerically exact
solution of the eigenvalue equation (\ref{eigex}) is restricted to small 
lattices, since the dimensionality of the spin space grows rapidly with the 
strip width $L$. 
Finite size results derived form transfer matrix calculations can be 
extrapolated only for a limited range of values of $L$.

The DMRG provides a very efficient algorithm for the construction of 
effective transfer matrices $\tilde{T}_L$ of large $L \times \infty$ strips,
which are generated iteratively from a transfer matrix of a strip of a small 
width ($L_0$) which can be diagonalized exactly.
In the initial step a row of the $L_0 \times \infty$ strip is divided into two
parts $A$ and $B$, which are labeled by variables $1 \leq j_A \leq N_A$ and
$1 \leq j_B \leq N_B$, such that $i\equiv (j_A,j_B)$ and $N_A N_B = Q^{L_0}$.
Typically we split the system into two subsystems of the same size, i.e. 
$N_A = N_B$, although this is not strictly necessary.

\begin{figure}[b]
\centerline{
\psfig{file=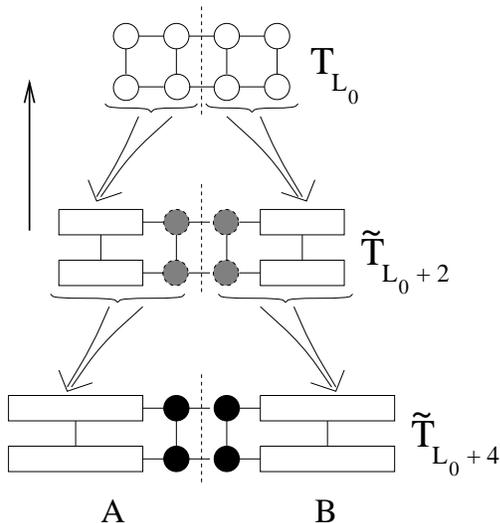,height=7cm}
}
\vskip 0.2truecm
\caption{Schematic view of the DMRG algorithm for the construction of effective 
transfer matrices $\tilde{T}_L$ (the arrow denotes the transfer direction). 
Starting from a small system ($L_0 =4$ in the example) the strip width is enlarged 
by two spins at each iteration. New spins are added at the center of the strip and 
rectangles denote blocks containing several spins, but labeled by $m$ states only.}
\label{FIG01}
\end{figure}

In the construction of a larger strip only part of the states of the $A$ and $B$
subsystems are kept, namely the eigenstates of the density matrices of $A$ and
$B$ corresponding to the largest eigenvalues.
The elements of the density matrix for the subsystem $A$ are defined as follows:
\be
\rho^{(A)} \left( l_A, k_A \right) =
\sum_{j_B = 1}^{N_B} 
v_{L_0}\left( l_A, j_B \right) \,\,
v_{L_0}\left( k_A, j_B \right)
\ee
where the sum is extended to all possible states of the subsystem $B$.
The density matrix eigenvalues $\omega_r$, which we order according to $\omega_1 > 
\omega_2 > \ldots > \omega_{N_A}$, are equal to the probabilities of finding the 
subsystem $A$ in the corresponding eigenvectors $| \Omega_r \rangle$ when the whole 
system is in a state $| v_{L_0} \rangle$.
In the calculation the spin space of the subsystem $A$ is truncated and one keeps 
$m$ eigenvectors $| \Omega_r \rangle$ corresponding to the largest eigenvalues. 
The truncation error, defined as:
\be
\epsilon = 1 - \sum_{r=1}^m \omega_r
\ee
gives an estimate of the accuracy of the procedure.
For $m = N_A$ (all states kept) obviously $\epsilon = 0$; typically the 
density matrix eigenvalues $\omega_r$ decrease rapidly with the index $r$, 
so good accuracy can be reached with a moderate number of states kept.
The truncated subsystem $A$ is enlarged by adding an extra Potts spin; 
the old subsystem plus the spin corresponds to a new subsystem $A$, labeled 
now by $m \, Q$ states.
An analogous procedure is followed for the subsystem $B$, whose spin space 
is first truncated by the selection of the dominant $m$ eigenvalues of the 
density matrix for the $B$ part $\rho^{(B)}$, and then enlarged by an extra spin.
Combining the two new subsystems $A$ and $B$ together one obtains a transfer 
matrix $\tilde{T}_{L_0+2}$ for a $(L_0 + 2) \times \infty$ strip.
The procedure is repeated until the strip of the wanted width has been 
generated:
\be
T_{L_0} \rightarrow
\tilde{T}_{L_0 + 2} \rightarrow
\tilde{T}_{L_0 + 4} \rightarrow \ldots \rightarrow
\tilde{T}_L.
\label{infinite}
\ee

\begin{figure}[b]
\centerline{
\psfig{file=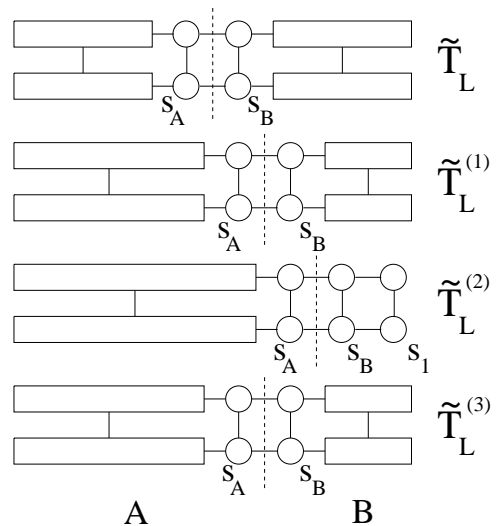,height=7cm}
}
\vskip 0.2truecm
\caption{Schematic view of White's finite system method. Only one of the two
subsystems grows at each iteration: subsystem $A$ in the first two iterations
and subsystem $B$ in the third. The spins $s_A$ and $s_B$ sweep trough the
strip width; from the transfer matrix $\tilde{T}_L^{(2)}$ one calculates, for
instance, the average magnetization of the surface spin $s_1$.}
\label{FIG02}
\end{figure}

A schematic view of the first iterations is shown in Fig. \ref{FIG01}.
In each DMRG step one solves the eigenvalue equation:
\be
\tilde{T}_L |\tilde{v}_L \rangle = \tilde{\lambda}_L |\tilde{v}_L \rangle
\ee
defined in a $Q^2 m^2 \times Q^2 m^2$ space. 
The quantity $\tilde{f}_L = - \ln(\tilde{\lambda}_L)/L$ approximate the 
exact free energy $f_L$ of the $L \times \infty$ strip and from the dominant 
eigenvector $| \tilde{v}_L \rangle$ one can calculate the magnetizations of 
the spins $s_{L/2,k}$ and $s_{L/2+1,k}$ at the center of the strip (see Fig. 
\ref{FIG01}) and the energy density of the bond connecting them.

The iterations in (\ref{infinite}) describe White's infinite system algorithm
\cite{whitePRB}. Repeating the procedure  many times one generates transfer 
matrices of very large systems: at each iterations new spins are added at the
center of the strip and the boundaries are pushed farther away from each other 
(see Fig. \ref{FIG01}).
Notice also that we implement open boundary conditions, as commonly done for DMRG
calculations since it has been found that the accuracy is the best in this case
\cite{whitePRB}.

For the calculation of the whole profile one must perform more DMRG iterations 
keeping the system size fixed:
\be
\tilde{T}_L \rightarrow \tilde{T}_L^{(1)} \rightarrow \tilde{T}_L^{(2)} 
\rightarrow \ldots \rightarrow \tilde{T}_L^{({\rm q})},
\label{finite}
\ee 
as shown schematically in Fig. \ref{FIG02} (we have taken $q \approx 4 L$
in our calculations).
In each of the iterations of (\ref{finite}) a spin is added only to one of 
the two subsystems, say to $A$, following the scheme described above. For 
the part $B$ one takes smaller subsystems generated in the previous DMRG 
iterations. This procedure is continued until the subsystem $B$ contains 
$L_0/2$ spins only, i.e. is known exactly.
Then in each of the following steps a spin is added to $B$ until the 
subsystem $A$ contains $L_0/2$ only.
In the present calculation this procedure has been repeated a couple of 
times.

In the iterations (\ref{finite}) the two spins $s_A$ and $s_B$ of 
Fig. \ref{FIG01} sweep through the whole strip width and from the sequence 
$|\tilde{v}_L \rangle$, $|\tilde{v}_L^{(1)} \rangle$, $|\tilde{v}_L^{(2)} 
\rangle$, \ldots one calculates the magnetization and energy density profiles 
across the strip. At the same time the free energies calculated from the
eigenvalues of the transfer matrices $\tilde{T}_L^{(i)}$ of (\ref{finite}) 
decrease at each iteration $\tilde{f}_L^{(i)} \leq \tilde{f}_L^{(i+1)}$ 
getting closer and closer to the exact free energy $f_L$ (it is known 
\cite{ostlund} that DMRG free energies provide upper bounds for $f_L$).

The steps in (\ref{finite}) describe the so-called finite system algorithm
\cite{whitePRB}, which is particularly indicated for the study of finite 
systems, since it provides higher accuracy than the infinite system method. 
The procedure is more time consuming, but has the advantage of providing 
information about the whole strip and it is necessary to calculate density 
profiles \cite{ourPRL}.
 
Using the finite system method we have generated effective transfer matrices 
for strips of widths $L = 8$, $16$, $24$ \ldots $64$ at the system critical 
point. We have used $m=32$ for $Q=2$, $m = 60$ for $Q=3$ and $m=80$ for $Q=4$; 
in each cases the truncation error was $\epsilon < 10^{-10}$.
It is not possible to relate the value of $\epsilon$ with the actual errors for 
the free energy and the density profiles. On the basis of our experience we 
expect an accuracy of $8-9$ digits for the density profiles for $L \leq 64$
and a somewhat higher accuracy for the free energy.

\section{Critical density profiles}
\label{sec:pro}

In a critical system confined between two parallel plates, being a large
but finite distance $L$ apart, the local densities $\langle \Phi(r) \rangle$
such as the magnetization and the energy density vary with the distance $l$ as a
smooth function of $l/L$. According to the scaling theory by Fisher and
de Gennes \cite{fishdg}
\be
\langle \Phi(l) \rangle_{ab} = l^{-x_{\Phi}} \,\, F_{ab}(l/L)\;,
\label{scal1}
\ee
where $ab$ denotes the boundary conditions (b.c.) applied at the two plates 
and $x_{\Phi}$ is the scaling dimension of the operator $\Phi$. 
For the magnetization and the energy density operators their value are 
connected with the more common critical exponents as:
\be
x_m = \beta/\nu \,\,\,\,\,\,{\rm and}\,\,\,\,\,\,\, x_\epsilon =d-1/\nu=
 (1 -\alpha)/\nu
\label{xmxe}
\ee
where $\beta$, $\nu$ and $\alpha$ are the magnetization, correlation length
and specific heat critical exponents, respectively, while $d$ denotes the
dimensionality of the system. 

\begin{figure}[b]
\centerline{
\psfig{file=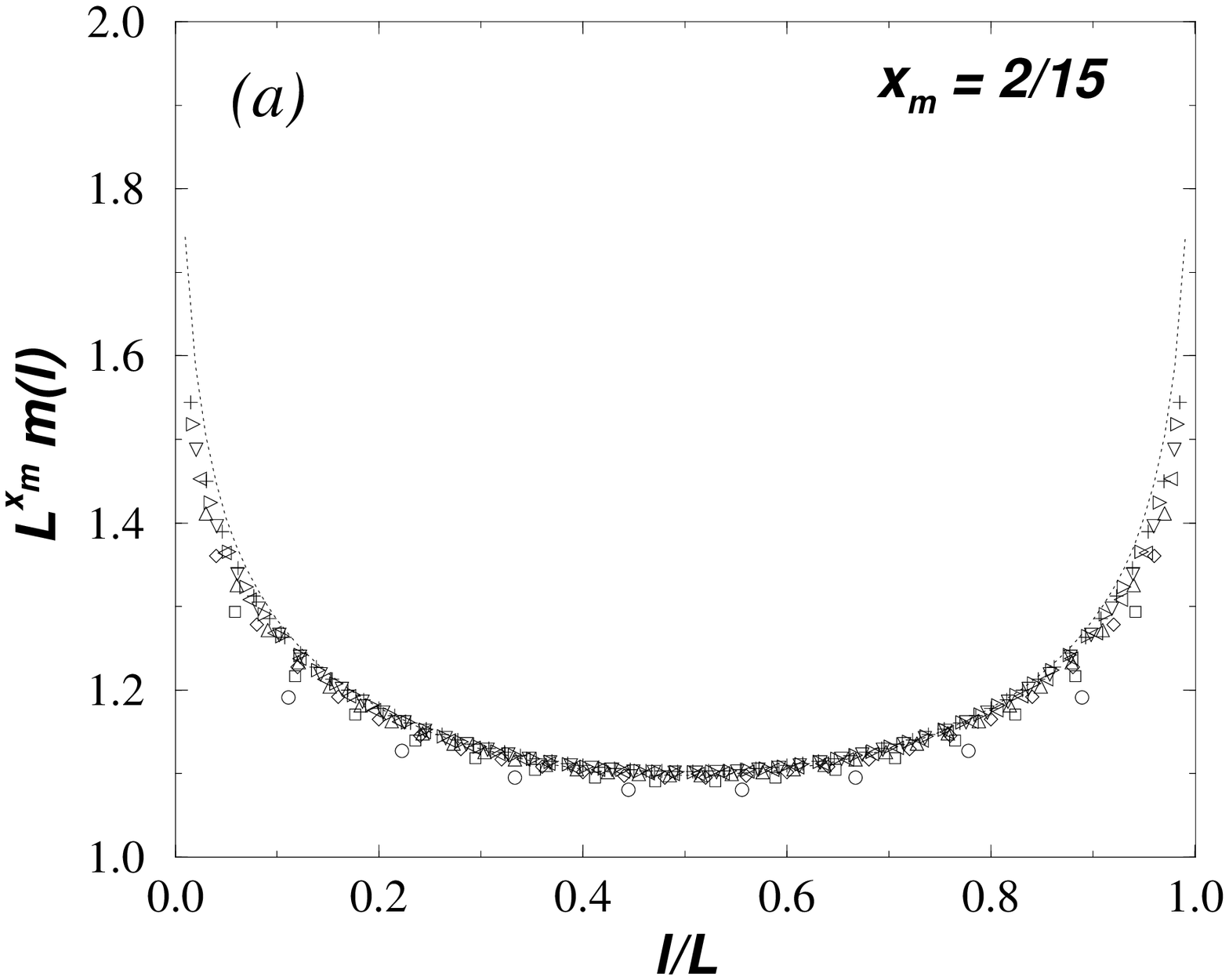,height=7.5cm}
}
\centerline{
\psfig{file=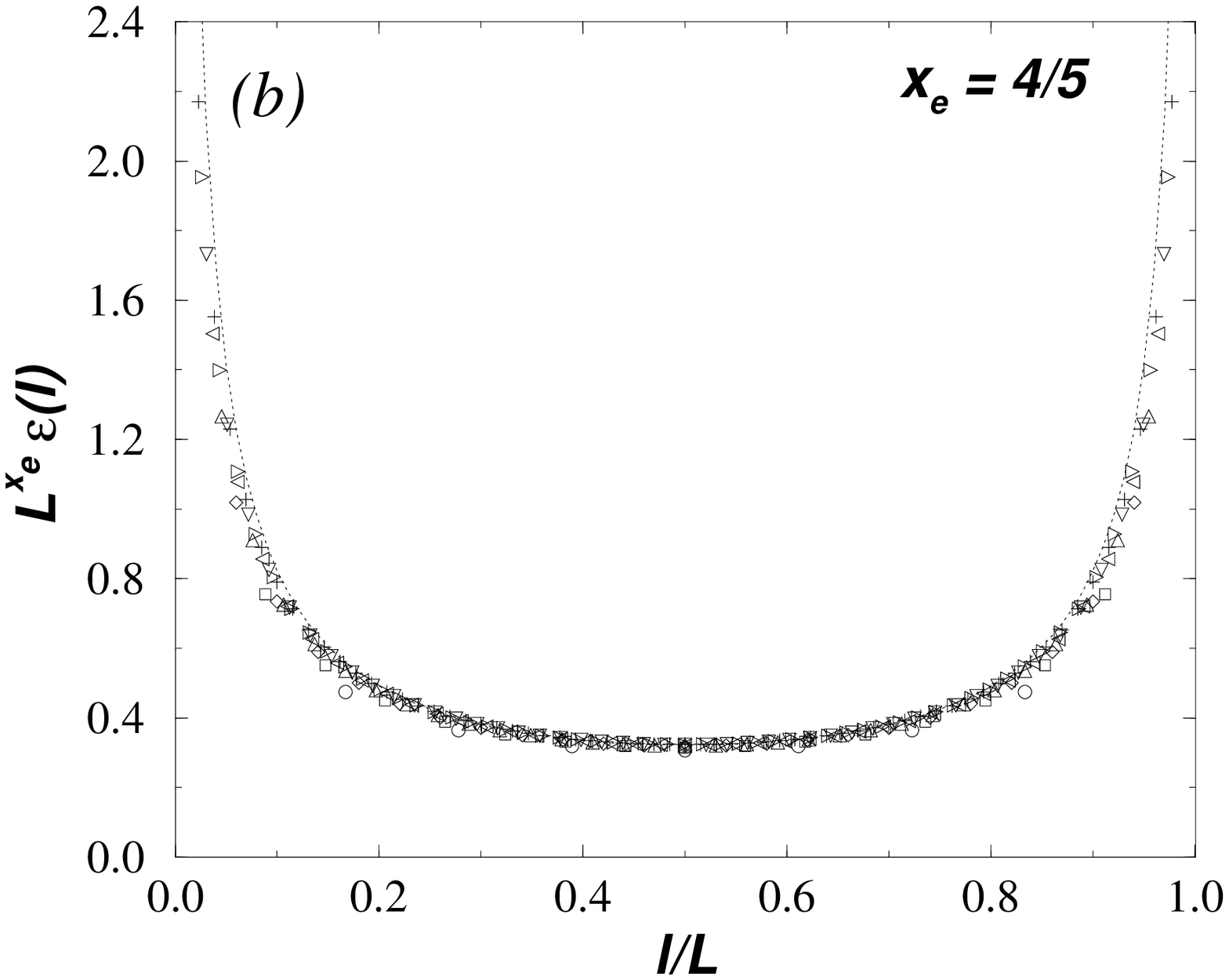,height=7.5cm}
}
\vskip 0.2truecm
\caption{Scaled magnetization (a) and energy density (b) profiles for 
the $Q = 3$ Potts model with fixed and equal spins at the boundaries. 
The symbols corresponds to $L \times \infty$ strips with $L=8$ (circles), 
$L=16$ (squares), $L=24$ (diamonds), $L=32$ (triangles up), $L=40$ 
(triangles down), $L=48$ (triangles left), $L=56$ (triangles right) and 
$L=64$ (pluses). The dotted lines are the conformal invariance predictions 
for the profiles. Also the values of the critical indices $x_m$ and $x_e$ 
used for the scaled profiles are indicated.}
\label{FIG03}
\end{figure}

The scaling function $F_{ab}$ in Eq. (\ref{scal1}) has the following
asymptotic behavior:
\be
F_{ab}(l/L) = {\cal A} \left[1+ B_{ab} \left( \frac{l}{L}\right)^d +
\ldots \right] \,\,\,\,\, {l \over L} \ll 1\;,
\label{scal2}
\ee
where $B_{ab}$, the Fisher-de Gennes coefficients are connected to 
universal quantities.

In two dimensions conformal invariance predicts the following form for the 
critical profiles with general conformally invariant b.c. \cite{burkxue,cardyprl}:
\be
\langle \Phi(l) \rangle_{ab}= \left[ \frac{L}{\pi} 
\sin \left( \frac{\pi l}{L} \right)
\right]^{-x_\phi} G_{ab} \left[ \cos \left( \frac{\pi l}{L} \right) 
\right]\;,
\label{confscal}
\ee
where the scaling function $G_{ab}(\omega)$ depends on the universality class of
the model and on the type of b.c. applied.
Expanding Eq. (\ref{confscal}) for $l/L \ll 1$ one finds that the Fisher-de Gennes
coefficients $B_{ab}$ in Eq. (\ref{scal2}) are given by:
\be
B_{ab} = \pi^2 \left[ \frac{x_\Phi}{6} - \frac{1}{2} \frac{{G_{ab}}'(1)}
{{G_{ab}}(1)} \right].
\ee

\begin{figure}[b]
\centerline{
\psfig{file=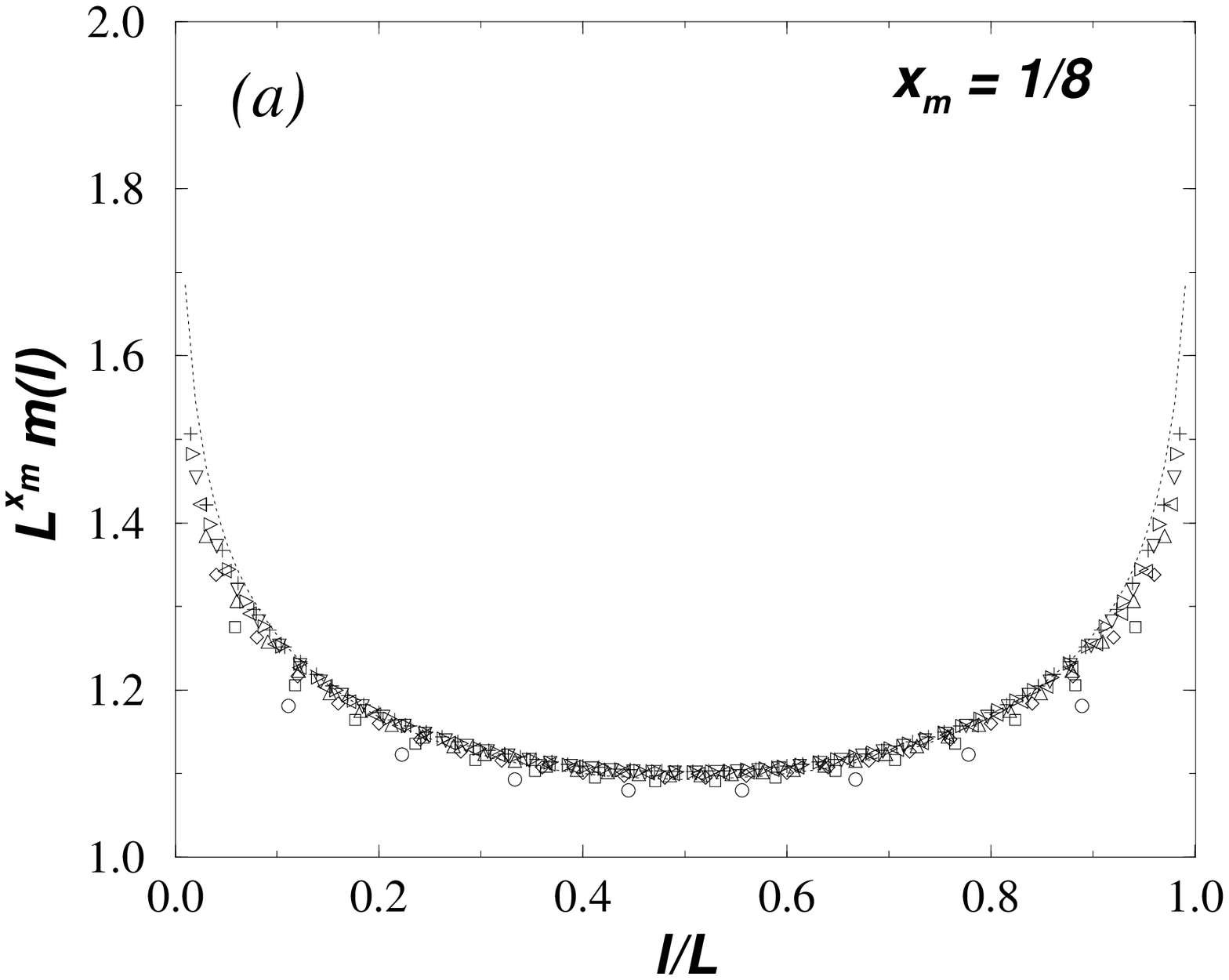,height=7.5cm}
}
\centerline{
\psfig{file=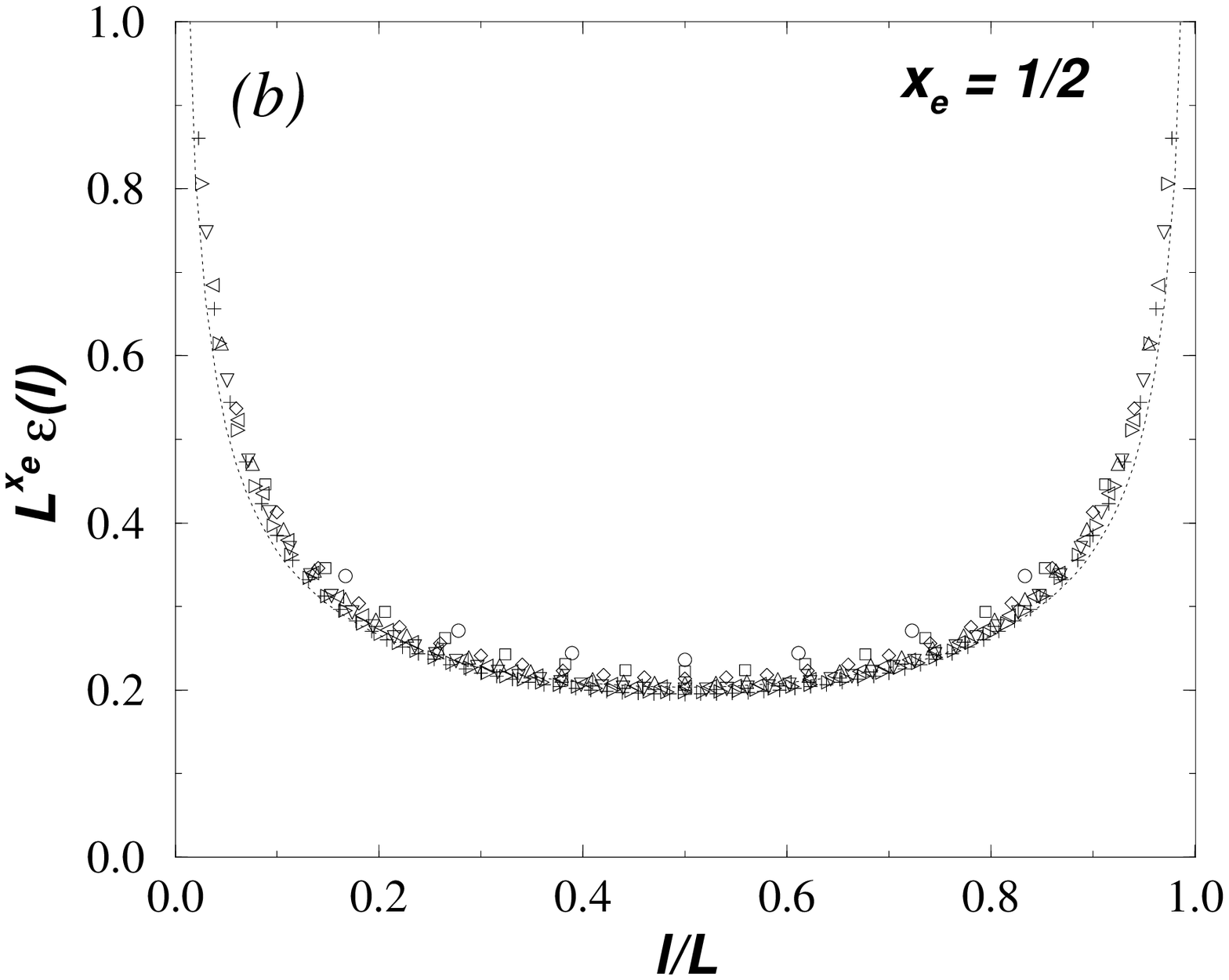,height=7.5cm}
}
\vskip 0.2truecm
\caption{As in Fig. \ref{FIG03}, but for $Q = 4$.}
\label{FIG04}
\end{figure}

In this section we present the critical profiles for the magnetization and 
the energy density calculated with the DMRG method and compare them with the
predictions of conformal invariance (\ref{confscal}). We have performed the
calculations for $Q=2$, $3$ and $4$ with all types of conformally invariant
b.c.
Here - for the sake of brevity - we present only the results
for $Q=3$ and $4$. The $Q=2$ (Ising) results, which can be calculated more
efficiently by the Pfaffian technique \cite{igloirieger97}, will be used in
the next Section to calculate the critical exponents of the model.
For all b.c. used we found that the $Q=2$ magnetization and energy 
density profiles agree well with the conformal results of 
Eq. (\ref{confscal}).

We start with the parallel spin b.c., i.e. we fix spins to the same state 
$s=0$ at the two boundaries and we indicate this choice by setting $a = b =0$ 
in Eq. (\ref{confscal}). According to conformal invariance 
\cite{burkhardteisenriegler} the scaling function in Eq. (\ref{confscal}) 
both for the magnetization and the energy density
is a constant: $G_{00}=const$. 
Notice that Eqs. (\ref{scal1}) and (\ref{confscal})
imply that the {\em scaled} density profile $L^{x_\phi} \langle \phi(l) 
\rangle_{ab}$ depends on $l$ only through the variable $l/L$.
In Figs. \ref{FIG03} and \ref{FIG04} we show the scaled magnetization
and energy density profiles for
the $Q=3$ and $Q=4$ models, respectively, together with the conformal results.
As seen in the Figures the DMRG data corresponding to different 
values of $L$
collapse into single scaling curves, which are in good agreement with
the conformal results. In the scaling plot we have used the values of the
scaling dimensions conjectured by conformal field theory and which are 
given in Table \ref{TAB03}. 
These scaling dimensions will be computed from a finite size scaling 
analysis of the DMRG data in the next Section.
Notice that the magnetization profiles of Fig. \ref{FIG03}(a) and  
Fig. \ref{FIG04}(a) differ only slightly since the magnetic exponents 
in the $Q=3$ and $Q=4$ Potts model are very close to each other.
A more visible difference between the $Q=3$ and the $Q=4$ case can 
be seen in the energy density profiles of Figs. \ref{FIG03}(b) and
\ref{FIG04}(b).

As next case we consider the fixed-free b.c., i.e. we fix the
spins to the state $s=0$ only at one boundary of the system; we indicate 
these b.c. by setting $a=0$, $b=f$ in Eq. (\ref{confscal}).
The scaling behavior of the critical densities in a free surface are
governed by the surface scaling dimensions \cite{binder}, which are 
generally different from the corresponding bulk quantities. For the 
$Q$ state Potts model the surface magnetization scaling dimensions
\be
x_m^s=\beta_s/\nu\:,
\label{xms}
\ee
where $\beta_s$
is the surface magnetization critical exponent, are listed in Table 
\ref{TAB03} as derived by conformal field theory \cite{cardy++}. 
We recall that the value of the surface energy scaling dimension is universally 
$x_e^s=2$ for two dimensional models \cite{burkhcardy}. 
The conformal prediction \cite{burkxue} about the scaling function in 
the critical profiles in Eq. (\ref{confscal}) is:
\be
G_{0f}^m = {\cal A} \left[ \cos \left( \frac{\pi l}{2 L} \right) \right]^{x_m^s}
\label{mag+f}
\ee
for the magnetization and
\be
G_{0f}^e = {\cal B} \cos \left(\frac{\pi  l}{L} \right)
\label{ene+f}
\ee
for the energy density (${\cal A}$ and ${\cal B}$ are constants).

\begin{figure}[b]
\centerline{
\psfig{file=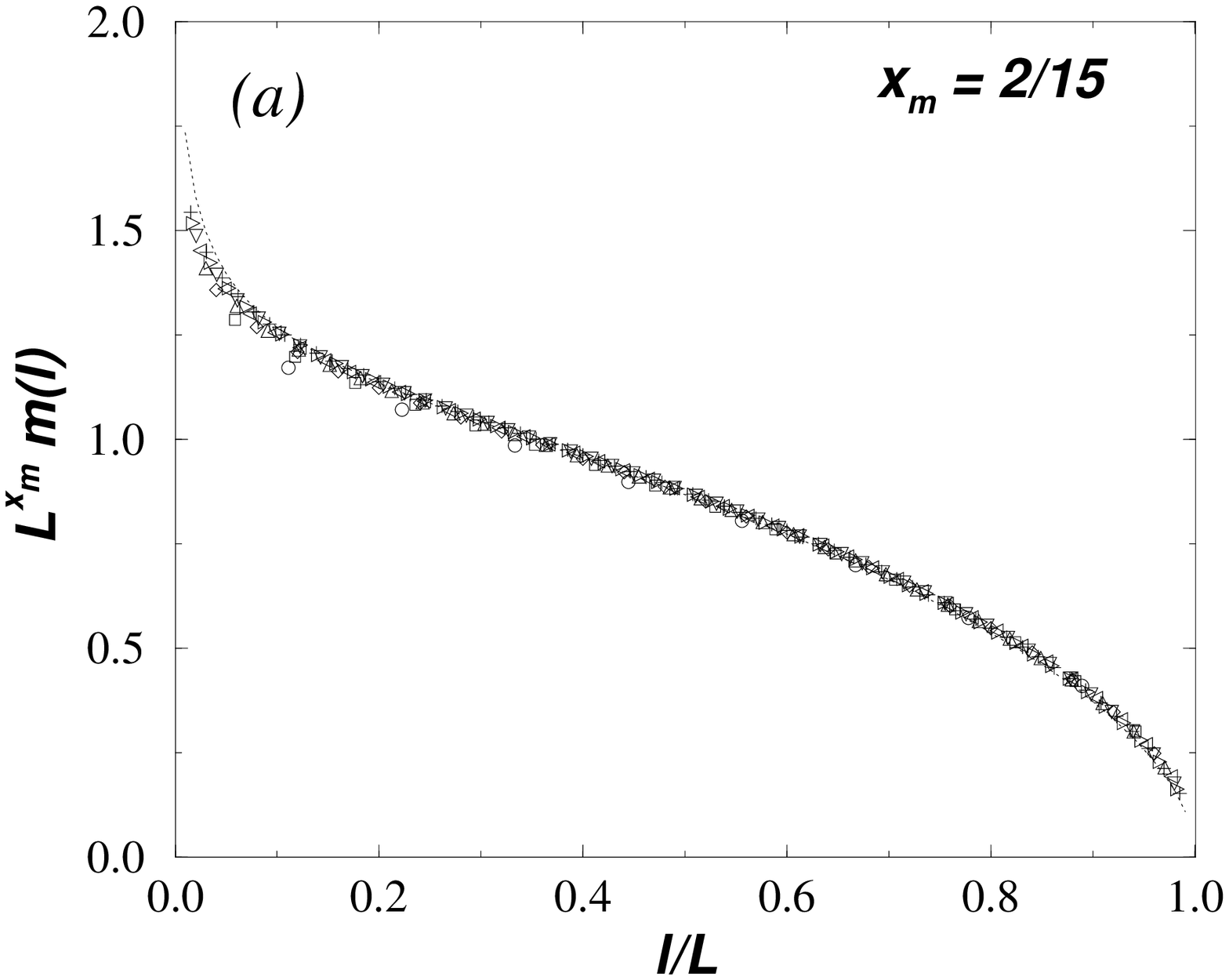,height=7.5cm}
}
\centerline{
\psfig{file=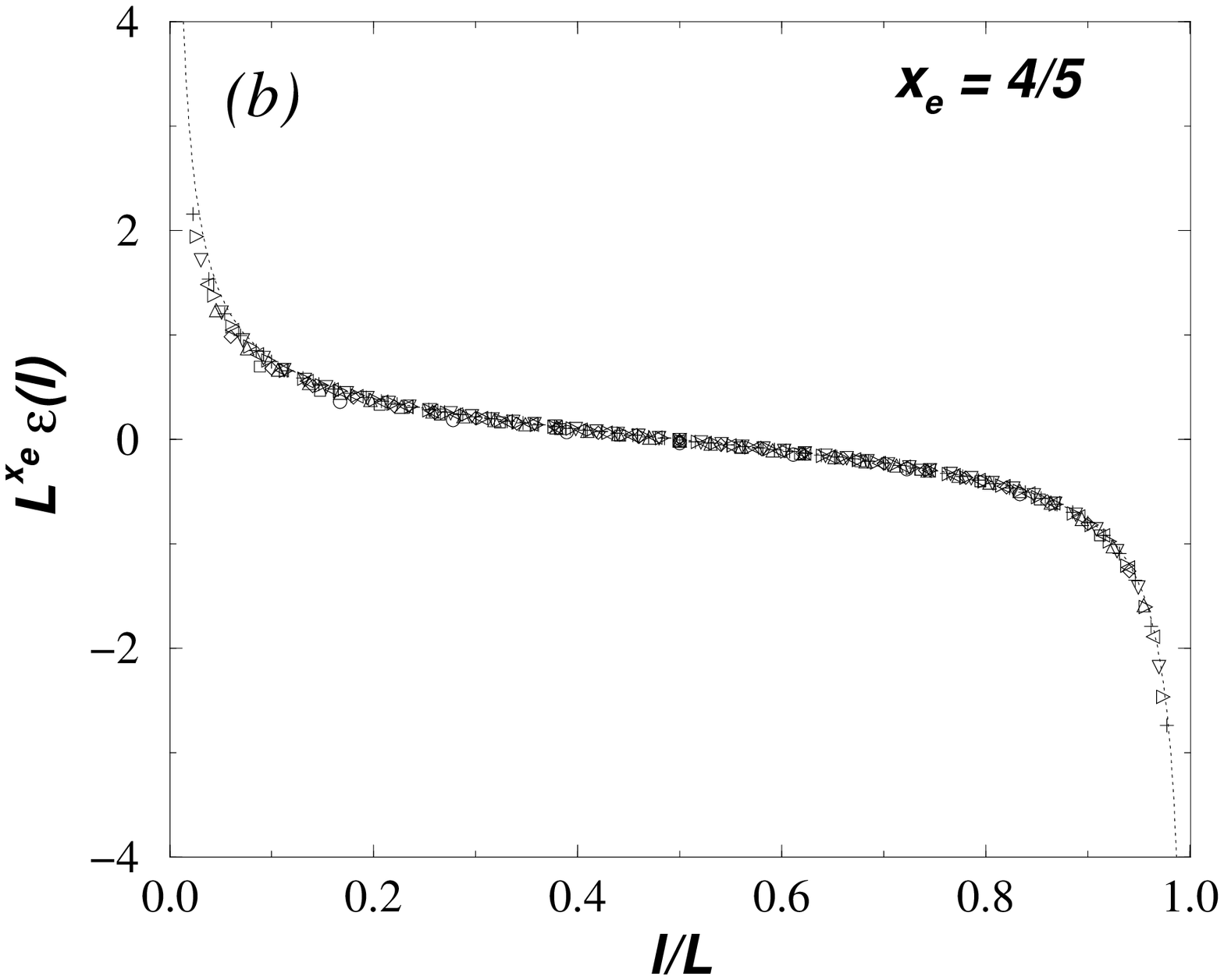,height=7.5cm}
}
\vskip 0.2truecm
\caption{Scaled magnetizations (a) and energy densities (b) for $Q=3$
in strips with fixed free boundary conditions. The symbols are as in
Fig. \ref{FIG03}.}
\label{FIG05}
\end{figure}

\begin{figure}[b]
\centerline{
\psfig{file=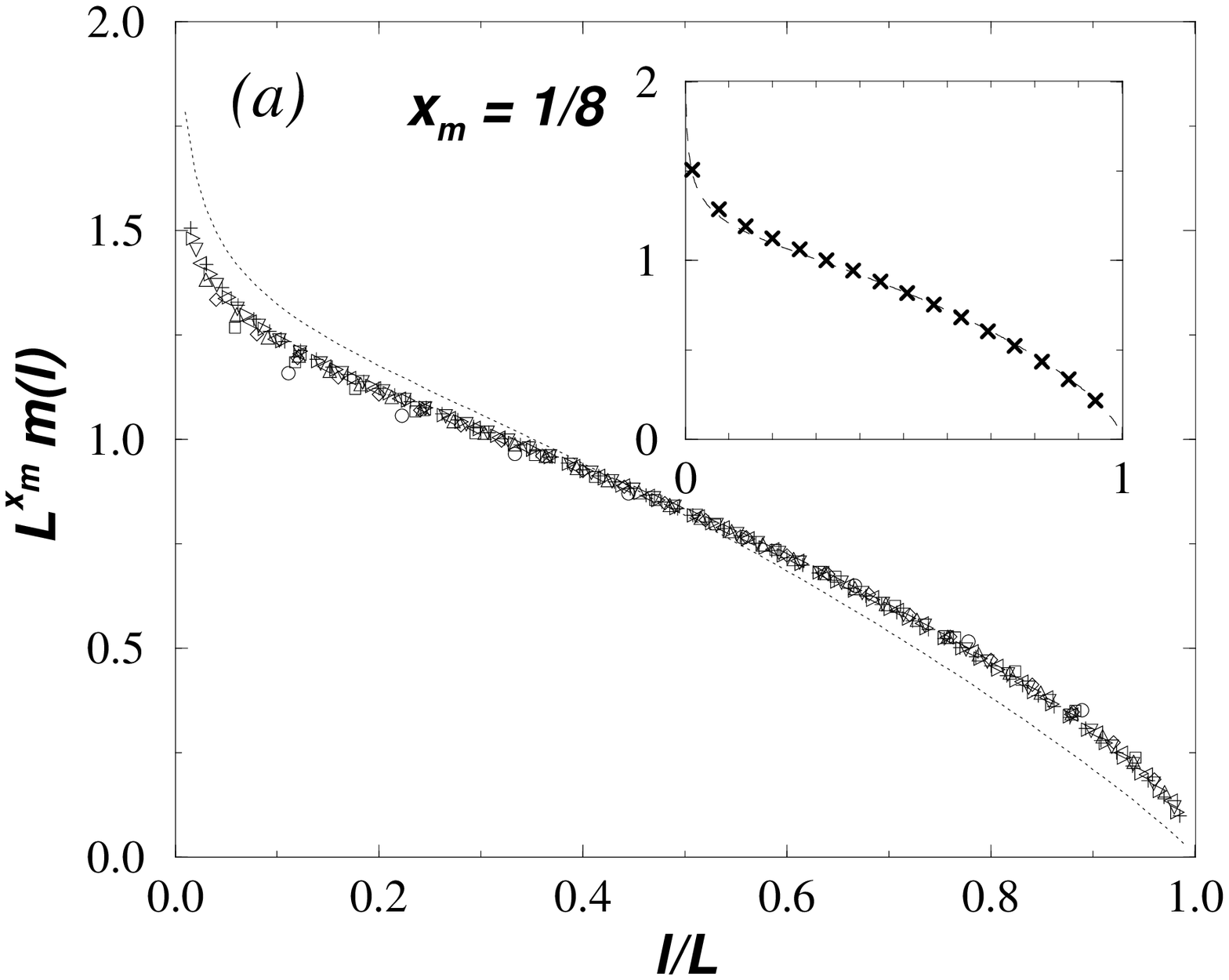,height=7.5cm}
}
\centerline{
\psfig{file=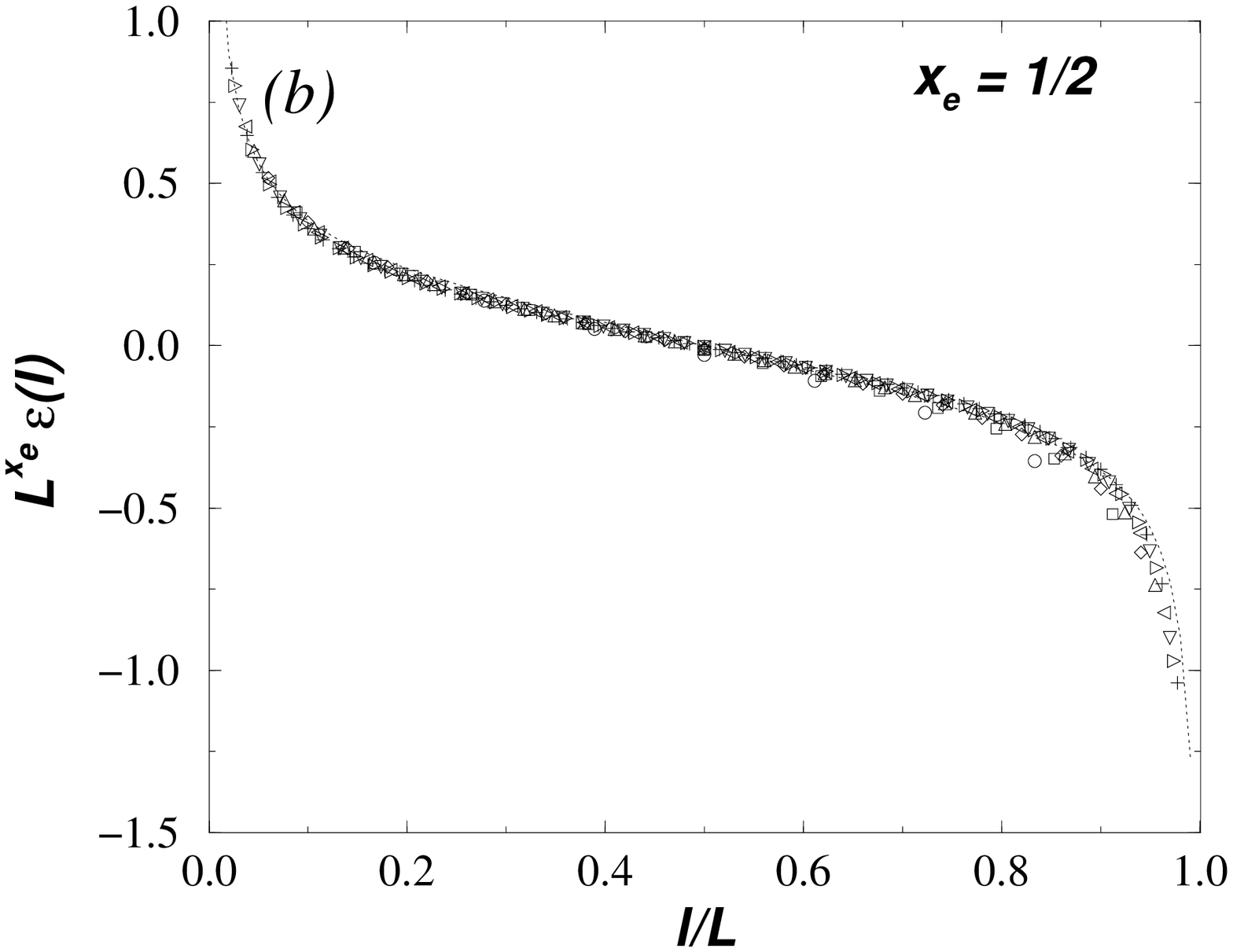,height=7.5cm}
}
\vskip 0.2truecm
\caption{As in Fig. \ref{FIG05} for $Q = 4$. The inset of (a) shows
the DMRG data for $L=64$ (crosses) and the conformal profile
of Eq. (\ref{confscal}), (\ref{mag+f}) where we have used effective
size dependent scaling dimensions $x_m(L)$ and $x_m^s(L)$ given in
(\ref{effexxm}) and (\ref{effexxms}) with $L = 64$.}
\label{FIG06}
\end{figure}

Figures \ref{FIG05} and \ref{FIG06} show the scaled magnetization and
energy density profiles with fixed-free b.c. for $Q=3$ and $Q=4$, respectively. 
The data collapse into single scaling curves which are in 
good agreement with the conformal invariance predictions (\ref{mag+f}) 
and (\ref{ene+f}), except for the magnetization profiles of the $Q=4$
state Potts model as shown in Fig. \ref{FIG06}(a) where one finds strong 
deviation of the numerical data from the conformal invariance results.
The origin of these discrepancies is the presence of very strong, 
logarithmic corrections to scaling in the $Q=4$ model \cite{logcorr}. 
As will be discussed in the next Section the leading $1/\ln L$ corrections 
to finite size scaling are universal and they can be taken into account 
using effective, size dependent scaling dimensions. 
We made use the same strategy here and for the largest finite system with 
$L=64$ we compared the DMRG data with the conformal results,
where in Eqs. (\ref{confscal}), (\ref{mag+f}) the corresponding 
effective scaling dimensions were used. 
As seen in the inset of Fig. \ref{FIG06}(a) we have obtained good
agreement with the corrected formula.

\begin{figure}[b]
\centerline{
\psfig{file=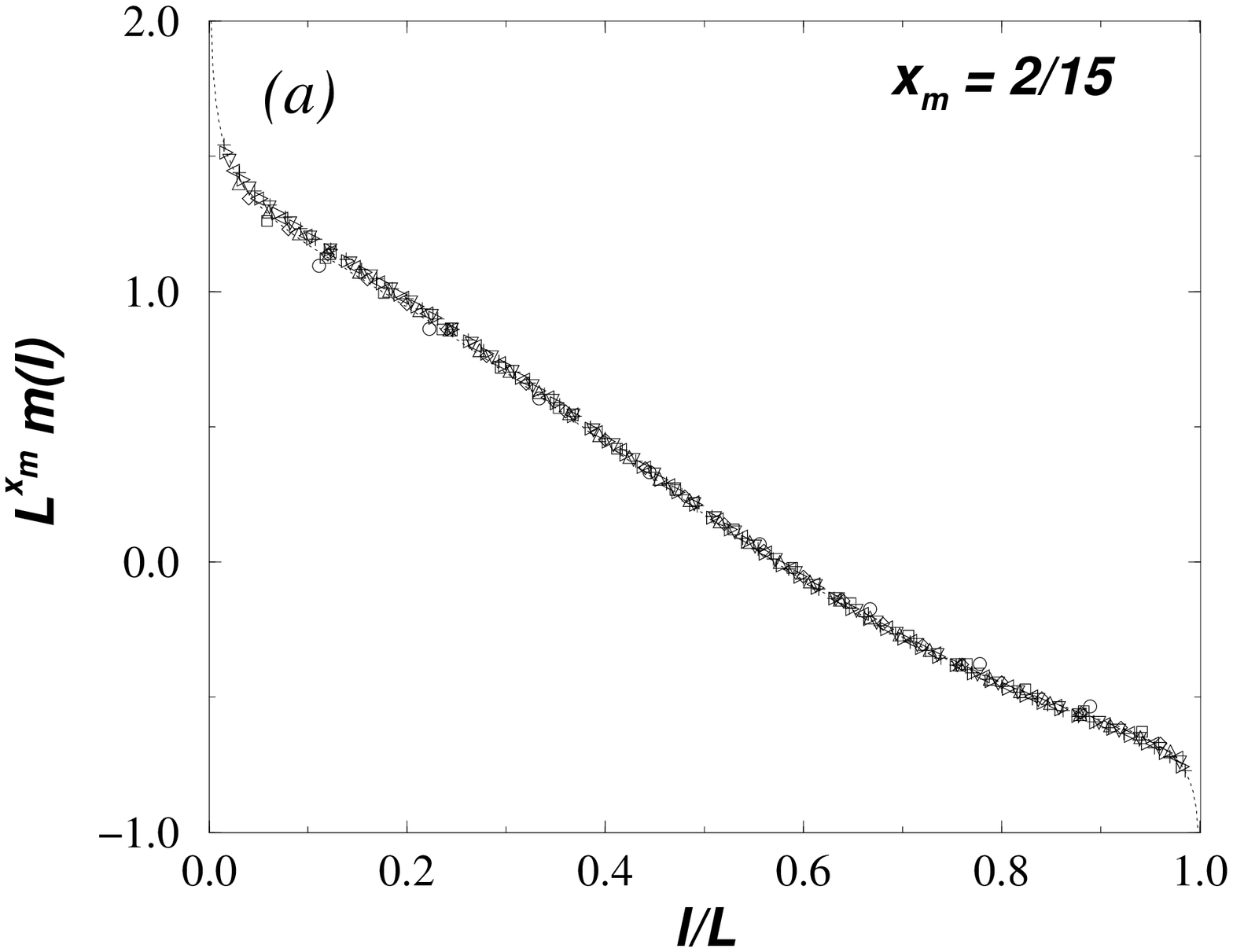,height=7.5cm}
}
\centerline{
\psfig{file=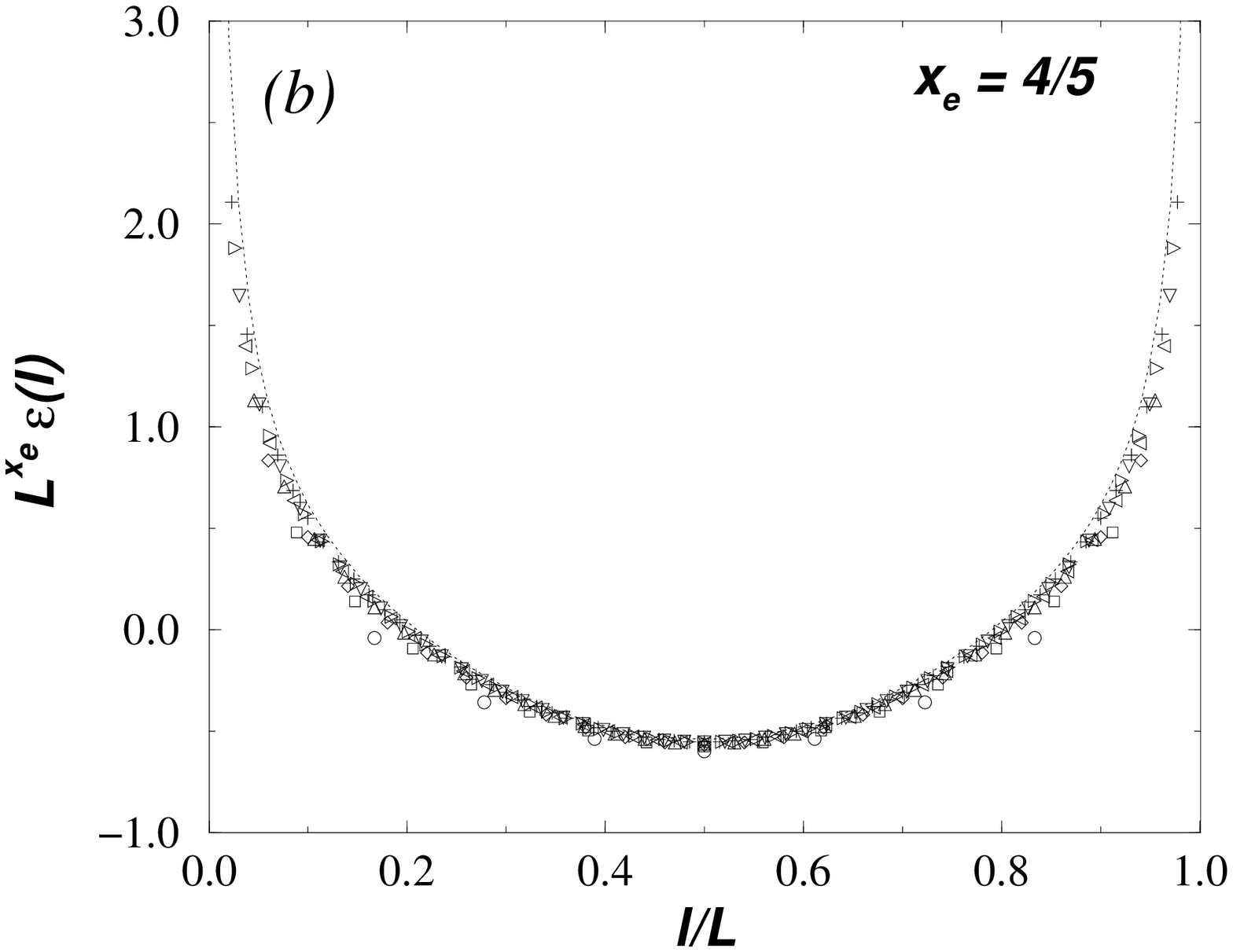,height=7.5cm}
}
\vskip 0.2truecm
\caption{Scaled magnetizations (a) and energy density profiles (b) in
strip with mixed boundary conditions for $Q=3$. The symbols are as in
Fig. \ref{FIG03}.}
\label{FIG07}
\end{figure}

Finally, we consider the third type of conformally invariant b.c., when 
the two boundary layers are fixed to different states, say $a=0$ and $b=1$. The
conformal predictions about the scaling functions in Eq. (\ref{confscal}) are
in this case the most complicated\cite{burkxue}. For the magnetization profile
the scaling function $G_{01}^m(\omega)$ is expressed by a combination of
hypergeometric functions, which reduces to polynomials for $Q=2$, where
$G_{01}^m(\omega)= {\cal A} \omega$ and for $Q=4$, where
$G_{01}^m(\omega)= {\cal B} (1+4 \omega + \omega^2)$. 
For the energy density profile the scaling function
\be
G_{01}^e = {\cal C} \left[ 1 - 8 \frac{1 - x_\epsilon/2}
{5 - 4 x_\epsilon} \sin^2 \left(\frac{\pi l}{L}\right) \right]\;
\label{ene+-}
\ee
involves the energy scaling dimension $x_\epsilon$ (${\cal A}$, ${\cal B}$ 
and ${\cal C}$ denote non-universal constants).

\begin{figure}[b]
\centerline{
\psfig{file=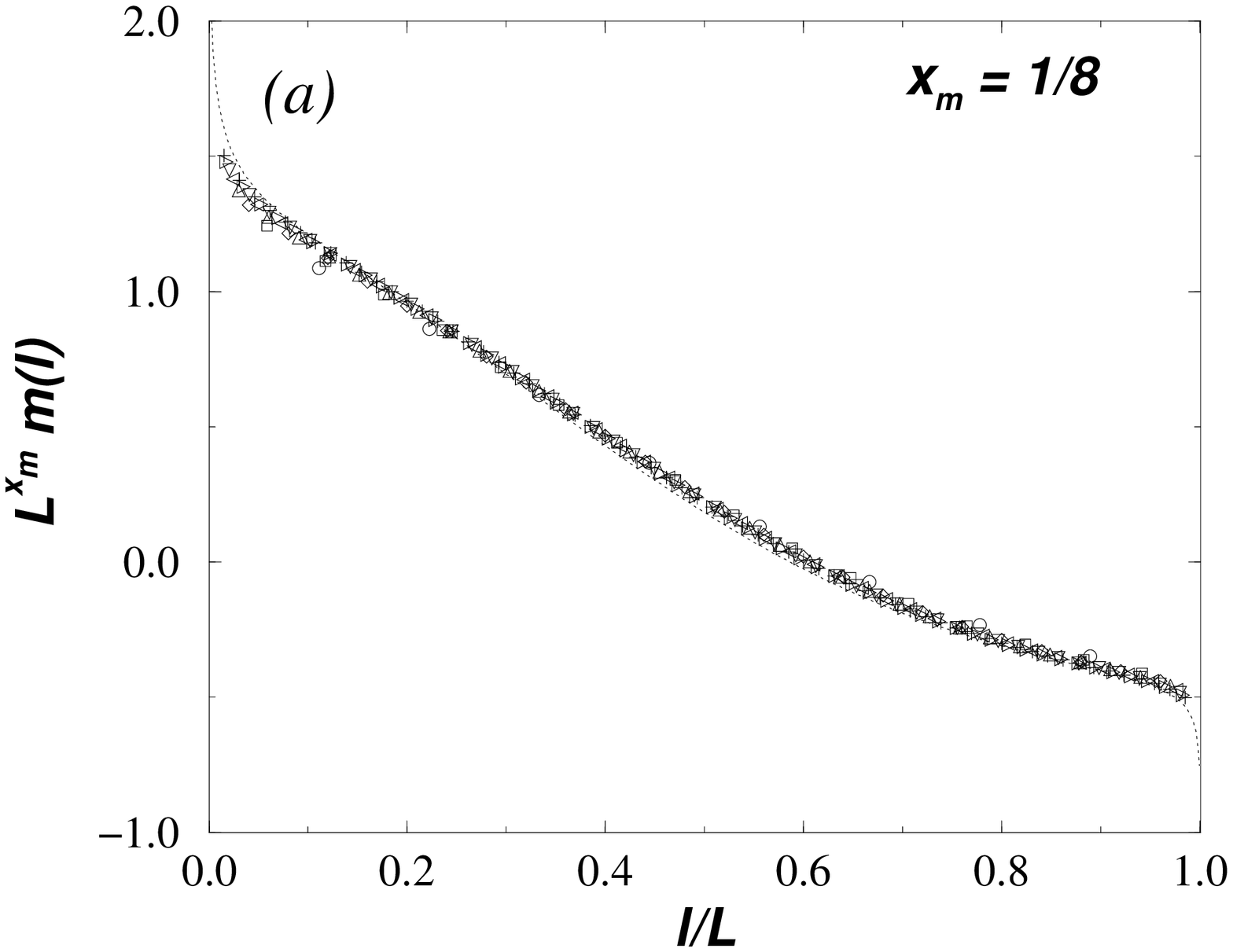,height=7.5cm}
}
\centerline{
\psfig{file=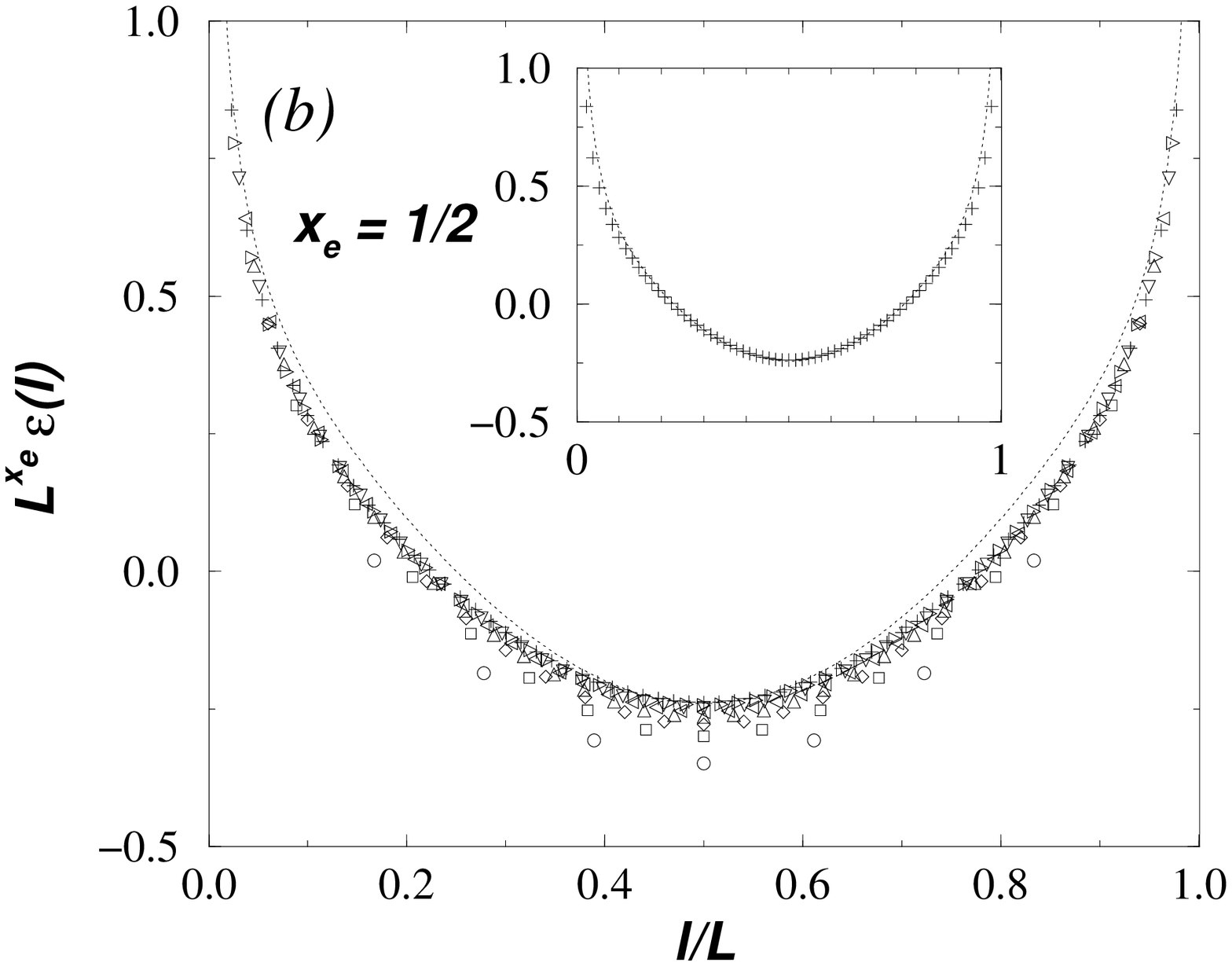,height=7.5cm}
}
\vskip 0.2truecm
\caption{As in Fig. \ref{FIG07} for $Q = 4$.
The inset of (b) shows the DMRG data for $L=64$ (pluses) and the profile 
obtained from conformal invariance where we have used an effective size 
dependent critical index $x_e (L)$ given in Eq. (\ref{effexxe}), for $L=64$.
}
\label{FIG08}
\end{figure}

Figures \ref{FIG07} and \ref{FIG08} show the scaled magnetization and energy 
density profiles for $Q = 3$ and $Q=4$ for mixed boundary conditions.
In the $Q=3$ case the numerical data referring to strips of different widths
collapse into single curves which are in very good agreement with the conformal 
invariance predictions. 
For the case $Q=4$ the agreement with conformal invariance is worse, again due
to the logarithmic corrections. The strongest corrections are found for the 
energy density profile Fig. \ref{FIG08}(b); the data collapse is clearly
worse than in the $Q=3$ case for the energy density.
As done for the fixed-free boundary conditions we consider finite size
approximants to the conformal profile (\ref{ene+-}), by substituting the
exponent $x_\epsilon$ appearing in Eq. (\ref{ene+-}) with $x_\epsilon(L)$
for $L=64$. Also in this case the modified conformal profile with an
effective size dependent exponent agrees quite well with the numerical 
data as seen in the inset of Fig. \ref{FIG08}(b).

\section{Critical exponents and Casimir amplitudes}
\label{sec:crit}

In this Section we perform a quantitative analysis of the finite lattice
data obtained by the DMRG method. First, we consider the various critical
exponents, then we investigate the finite size dependence of the free-energy
density and calculate the Casimir amplitudes with different boundary conditions.

\subsection{Critical exponents}

The critical exponents such as $x_m$, $x_e$ and $x_m^s$ defined in
Eqs. (\ref{xmxe}) and (\ref{xms})
can be calculated by two different methods. In the first method, which is
more general and not restricted to two dimensional systems, we use the
{\em scaling form} of Eq. (\ref{scal1}).
The second way of determination of the critical exponents is based on 
{\it conformal invariance}: we use explicitly the functional form of the 
density profiles presented in Sec. \ref{sec:pro}.

We start with the scaling method and compare the values of a density, 
for instance the magnetization $m_L(l)$ for different widths $L$,
which are taken as $L=8$, $16$,\dots $64$, and at constant values of 
$z = l/L$. 
The corresponding scaling dimension - $x_m$ - is obtained as the limiting 
value of the finite lattice estimates:

\be
x_m(L)=\frac{\ln\left(m_L/m_{L+8}\right)}{\ln(1+8/L)}\;,
\label{fssexp}
\ee
which - according to scaling theory - is independent of $0<z<1$
(in the previous formula $m_L \equiv m_L(z L)$ and $m_{L+8} \equiv 
m_{L+8}[z (L+8)]$). 

Following traditional finite size scaling methods \cite{barber} we extrapolate the 
finite lattice approximants $x_m(L)$ to $L \to \infty$, using a powerful sequence 
extrapolation technique, such as the widely used BST-method \cite{BST}. Here, we are 
not going to recapitulate the details of this extrapolation procedure just for 
illustration we present the table of extrapolants (Table \ref{TAB01}) for $x_m$ 
obtained at the middle point ($z=1/2$) of the magnetization profiles of the $Q=3$ 
Potts model with parallel-spin boundary conditions. From the original series, which 
is in the first column of Table \ref{TAB01} and calculated through Eq. (\ref{fssexp}),
new extrapolants are generated by the repeated use of the BST algorithm,
which are shown in the successive columns. 
One generally expects faster convergence within the data
in higher order columns. As one can see in Table \ref{TAB01} there is a nice 
convergence up to the third and fourth column, which however is not increased 
in the further steps. Thus our estimate of the magnetic scaling dimension 
based on the BST algorithm is
\be
x_m = 0.1334(1)~~~~~~(Q=3)\;,
\label{xmq3}
\ee
which agrees very well with the conformal result $x_m =2/15$ \cite{dotsenko,cardy}, 
and it is at least as accurate as the other existing numerical results\cite{bloete}. 

\end{multicols}\widetext
\begin{table}[hb]
\caption{Table of the BST extrapolants for the bulk magnetization exponent for the 
$Q=3$ Potts model calculated from finite size scaling at the center of the strip 
with fixed identical spins at the boundaries.
(The parameter of the BST algorithm
is $\epsilon=5/6$.)
}
\label{TAB01}
\begin{tabular}{cccccccc}
\vspace{.5mm}\\
0.108817 & 0.128823 & 0.133723 & 0.133518 & 0.133505 & 0.133511 & 0.133715  \\
0.116774 & 0.131197 & 0.133577 & 0.133465 & 0.133510 & 0.133801 & \\
0.120575 & 0.132097 & 0.133514 & 0.133431 & 0.133603 & & \\           
0.122859 & 0.132535 & 0.133475 & 0.133409 & & & & \\         
0.124400 & 0.132781 & 0.133449 & & & & \\      
0.125517 & 0.132933 & & & & & \\
0.126367 & & & & & & \\
\vspace{.5mm}\\
\end{tabular}
\end{table}
\begin{multicols}{2}\narrowtext

For the calculation of the scaling dimension $x_\epsilon$ we have used an expression
similar to Eq. (\ref{fssexp}) with the magnetization replaced by the energy density 
profile. The surface exponent $x_m^s$ follows from the analysis of the magnetization 
profile for fixed-free boundary conditions; according to scaling theory the magnetization
of the surface spin at criticality in a $L \times \infty$ strip scales as:
\be
m_L^s \sim L^{-x_m^s}
\label{surfscal}
\ee
from which one can extract the surface exponent in an analogous way as done for 
$x_m$ (see Eq. (\ref{fssexp})).

Next, we are going to describe our second way of calculating the critical exponents,
which is based on the conformal results about the profiles. As an example we consider the
magnetization profile with fixed-free b.c. in Eqs. (\ref{confscal}), (\ref{mag+f})
and take three points from the profile for a strip of given width $L$: 
$m_L(L/4)$, $m_L(L/2)$ and $m_L(3 L/4)$. 
It is easy to see that the surface magnetization exponent $x_m^s$ is approximated as:

\be
x_m^s(L)={\ln\left[m_L\left(\frac{L}{4}\right)/m_L\left(\frac{3L}{4}
\right)\right]\over \ln(\sqrt{2}+1)}\;,
\label{confexps}
\ee
whereas the bulk exponent $x_m$ is involved in the combination:

\be
x_m(L)- \frac{x_m^s(L)}{2}
={\ln\left[m_L\left(\frac{L}{4}\right) m_L\left(\frac{3L}{4}\right) 
/ m_L\left(\frac{L}{2}\right)^2\right]\over \ln 2 }\;.
\label{confexp}
\ee

\end{multicols}\widetext
\begin{table}[hb]
\caption{Table of the BST extrapolants for the surface magnetization exponent
for the $Q=3$ Potts model calculated from Eq. (\ref{confexps}).
(The parameter of the BST algorithm is $\epsilon=5/6$.)
}
\label{TAB02}
\begin{tabular}{ccccccc}
\vspace{.5mm}\\
0.554998 & 0.673224 & 0.668052 & 0.667014 & 0.666865 & 0.666938 & 0.666666 \\
0.584425 & 0.670860 & 0.667623 & 0.666936 & 0.666799 & 0.666574 & \\  
0.600929 & 0.669696 & 0.667388 & 0.666880 & 0.666762 & & \\ 
0.611585 & 0.669012 & 0.667238 & 0.666839 & & & \\ 
0.619076 & 0.668565 & 0.667136 & & & & \\
0.624650 & 0.668252 & & & & & \\ 
0.628972 & & & & & & \\
\vspace{.5mm}\\
\end{tabular}
\end{table}
\begin{multicols}{2}\narrowtext

Using other profiles one can easily obtain relations for the bulk (magnetization
or energy) exponents, as well. To illustrate the accuracy of the conformal method
we show in Table \ref{TAB02} the BST extrapolants of the surface magnetization
exponent, calculated from Eq. (\ref{confexps}) for the $Q=3$ model. Here, one can
see a similar tendency in the extrapolations as in Table \ref{TAB01}
for the finite size data. The obtained estimate of the surface exponent 
\be
x_m^s = 0.6667(1)~~~~~~~(Q=3)\;,
\label{xmsq3}
\ee
is again in excellent agreement with Cardy's conformal result
$x_m^s=2/3$ \cite{cardy++}.
We note that our estimates on the critical exponents for a given model,
obtained by the scaling and the
conformal methods are consistent and the errors of the two methods are also
comparable. The same is true if other type of profiles or other points of the
profiles are considered. We used this freedom to obtain an objective criterion
about the error of our estimates.

Our results for the critical exponents calculated by the DMRG method are presented
in Table \ref{TAB03} together with the exact and conformal results. The
estimates are most accurate for the $Q=2$ Ising model, which
has two reasons. First, the structure of confluent singularities are the simplest
for this model, and second, the numerical accuracy of the DMRG method is also
the most accurate in this case. The estimates for the $Q=3$ model are still
very accurate, as a matter of fact the accuracy of the DMRG method in this case
is comparable or even higher than other numerical methods (Monte Carlo
simulation \cite{MC}, series expansion \cite{series}, traditional finite size
scaling\cite{bloete}, corner transfer matrix approach\cite{cornerTM,loic},etc).

The $Q=4$ state model, for which the estimates in Table \ref{TAB03} are less
accurate, needs special considerations. It is known from a RG analysis that the
critical Hamiltonian of the model involves a marginal scaling field $\Psi$, which
results in logarithmic corrections to scaling. Then the finite-size scaling behavior
of the free-energy density of the model as a function of the bulk ($t,h$)
and surface ($h_s$) scaling fields is given by \cite{logcorr,attilio}:
\end{multicols}\widetext
\be
f(t,h,h_s,\Psi,L) =
L^{-d}f(L^{2-x_e}Z^{3/4}t,L^{2-x_m}Z^{1/16}h,L^{1-x_m^s}Z^{-1}h_s,Z\Psi,1)\;,
\label{freelog}
\ee
\begin{multicols}{2}\narrowtext
with
\be
Z=\left(1-{\Psi(0)\over \pi} \ln L \right)^{-1}\;.
\label{Z}
\ee
Now the different physical quantities can be obtained through differentiation
and then their singular behavior at the critical point can be studied by finite
size scaling. It is easy to see from Eq. (\ref{freelog}) that the critical exponents
determined by this way have universal $1/\ln L$ corrections, which does not depend
on $\Psi(0)$:
\be
x_m(L)={1 \over 8} + {1 \over 16}{1 \over \ln L} + O[(\ln L)^{-2}]\;
\label{effexxm}
\ee

\be
x_\epsilon(L)={1 \over 2} + {3 \over 4}{1 \over \ln L} + O[(\ln L)^{-2}]\;
\label{effexxe}
\ee

\be
x_m^s(L)=1-{1 \over \ln L} + O[(\ln L)^{-2}]\;
\label{effexxms}
\ee

Then, the appropriate strategy to analyze the finite size estimates for 
the exponents \cite{bloete1}, such as for the bulk magnetization scaling 
dimension $x_m$, is to consider the series of effective exponents 
$x_m(L)-1/(16 \ln L)$, which has then only $(1/\ln L)^{2}$ corrections. 
The results of this type of analysis are presented in Table \ref{TAB03}, 
which are in satisfactory agreement with the conformal results and also 
give support to the validity of the RG analysis leading
to Eq. (\ref{freelog}). 

\end{multicols}\widetext
\begin{table}[hb]
\caption{Comparison between exact ($\dagger$) or conjectured $(\ddagger)$
bulk and surface exponents for the $Q \leq 4$ Potts model and the numerical 
results obtained from finite size scaling extrapolations of DMRG data.}
\label{TAB03}
\begin{tabular}{ccccccc}
\vspace{1mm}\\
Q & $x_m$  & $x_m$ (DMRG) & $x_{m_s}$  & $x_{m_s}$ (DMRG) 
& $x_\epsilon$  & $x_\epsilon$ (DMRG)  \\
\vspace{1mm}\\
\tableline
\vspace{1mm}
2 &  1/8$^{\dagger}$ & 0.125000(2)  & 1/2$^{\dagger}$  & 0.500000(5) 
& 1$^{\dagger}$  & 1.00000(1) \\
\vspace{.5mm}
3 &  2/15$^{\ddagger}$  & 0.1334(1)  & 2/3$^{\ddagger}$  & 0.6667(1) 
& 4/5$^{\ddagger}$ & 0.800(1) \\
\vspace{.5mm}
4 &  1/8$^{\ddagger}$ & 0.120(5)  & 1$^{\ddagger}$ & 1.02(2) 
& 1/2$^{\ddagger}$ & 0.49(2)\\
\end{tabular}
\end{table}
\begin{multicols}{2}\narrowtext

\end{multicols}\widetext
\begin{table}[hb]
\caption{Comparison of Casimir amplitude predicted by conformal invariance (CI)
and the numerical values obtained from extrapolations of the numerical data 
(DMRG) for different types of boundary conditions.}
\label{TAB04}
\begin{tabular}{llll}
\vspace{.5mm}\\
 & Q=2 & Q=3 & Q=4 \\
\vspace{.5mm}\\
\tableline
\vspace{.5mm}
\\
$A_{00}$ (CI)   & -0.0654498 $\left(-\frac{\pi}{48}\right)$ 
& -0.10472 $\left(-\frac{\pi}{30}\right)$ 
& -0.1309 $\left(-\frac{\pi}{24}\right)$ \\
$A_{00}$ (DMRG) & -0.065447(5) & -0.104(1) & -0.127(5) \\
$A_{0f}$ (CI)   & 0.13089969 $\left(\frac{\pi}{24}\right)$ 
& 0.287979 $\left(\frac{11 \pi}{120}\right)$
& 0.6545 $\left(\frac{5 \pi}{24}\right)$ \\
$A_{0f}$ (DMRG) &  0.1309003(5) & 0.2865(10) & 0.67(3) \\
$A_{01}$ (CI) & 1.5053465 $\left(\frac{23\pi}{48}\right)$
& 1.989675 $\left(\frac{19 \pi}{30}\right)$ 
& 3.0107 $\left(\frac{23 \pi}{24} \right)$ \\
$A_{01}$ (DMRG) & 1.505350(5) & 1.987(5) & 2.93(10) \\
\end{tabular}
\end{table}
\begin{multicols}{2}\narrowtext

\subsection{Casimir Amplitudes}

In the strip geometry with general conformally invariant boundary conditions
the finite size dependence of the critical free energy-density is given as:

\be
f_{ab}(L)=f_0+\frac{f_a+f_b}{L}+ \frac{A_{ab}}{L^2} + \dots\;,
\label{casimir}
\ee
where the bulk - $f_0$ - and the surface - $f_a+f_b$ - contributions are non-universal.
The second-order correction term $A_{ab}$, known as Casimir amplitude, involves 
universal quantities. As shown in Refs. \cite{cardyprl,burkxue}, the Casimir amplitude 
and the Fisher-de Gennes parameter $B_{ab}$ in Eq. (\ref{scal2}) are related as
\be
A_{ab}=-{c \over 4 \pi x_{\Phi}} B_{ab}^{\Phi}\;,
\label{fishercasimir}
\ee
where $c$ denotes the conformal anomaly number or central charge. 
We recall that $c = 1/2$ for $Q = 2$, $c = 4/5$ for $Q = 3$ and $c = 1$
for $Q = 4$.
From Eq. (\ref{fishercasimir}) combined with the results about the profiles in 
Section \ref{sec:pro} one obtains the values of the Casimir amplitudes predicted 
by conformal invariance. The simplest case is that of equal spins at the boundary 
for which the Casimir amplitude is just proportional to the conformal anomaly:
\be
A_{00}=-{\pi \over 24} 
\,\, c\;.
\label{11bc}
\ee
In the numerical calculation we made use of Baxter's exact result about the bulk 
free energy-density \cite{baxterpotts} for the $Q$ state Potts model
which is given by the following formula:
\be
f_0
={\ln Q \over 2} 
+\int_{-\infty}^{\infty} {dx \over x} \,\, {\rm tanh}(\mu x)
\,\, {\sinh (\pi-\mu)x \over \sinh(\pi x)}\;,
\label{f0}
\ee
where $\cos \mu = \sqrt{Q}/2$.

Having subtracted $f_0$ from $f_{ab}(L)$ we extrapolated the surface contributions 
and then the Casimir ampltudes using the BST algorithm. The results obtained for
the Casimir amplitudes for different b.c.  are presented in Table \ref{TAB04} 
together with the conformal results. 
As for the critical exponents the results for the Ising model ($Q=2$) are the 
most accurate, but good agreement between DMRG data and conformal results is 
found also for $Q=3$. We stress that the numerical values for the amplitudes
are obtained from two successive BST extrapolations.
In the case $Q=4$ also the Casimir amplitudes are affected by logarithmic 
corrections.
From the analysis of the extrapolants for the conformal anomaly, which can
be obtained numerically from Eq. (\ref{11bc}), we found that such quantity
has logarithmic corrections of type $1/(\ln L)^2$. There are no theoretical
predictions available about the form of these corrections. In the case of 
periodic b.c.  it is instead known \cite{cardylog}, from conformal results, 
that $c$ has corrections of the type $1/(\ln L)^3$.

\section{Discussion}
\label{sec:disc}

In this paper we have used the DMRG method to study the critical properties
of the two dimensional $Q$ state Potts model. Our study has demonstrated once 
again that, contrary to the widespread believe, the DMRG technique is able to 
describe accurately the critical region of a two dimensional classical (and 
consequently also of a one dimensional quantum) system.
Indeed the numerical accuracy of the critical exponents calculated by our approach 
is comparable or even better than those of the other available numerical methods.

As was noticed in early studies the DMRG technique faces into problems to
describe the critical region, in the thermodynamic limit of the infinite system method.
In this case, as was shown by \"Ostlund and Rommer \cite{ostlund}, the DMRG ground
state is of a tensor product type, which 
can not describe algebraically
decaying correlation functions, i.e. critical states. In the approach we
used we have escaped this problem, since first, we applied the
{\it finite system algorithm} 
and we restricted ourselves to lattices of moderate witdh,
and second, we applied {\it symmetry breaking boundary conditions}. 
In the finite systems studied the correlation length remains always finite
(and the transfer matrix spectrum is gapped), and the advantage of the boundary
conditions is that one can calculate critical exponents from the non-vanishing
value of the magnetization and energy density profiles. In the case of fixed-free
boundary conditions from the same profile one can derive surface and bulk exponents.

To analyze the critical point data we borrowed the method of the traditional
finite size scaling and used efficient sequence extrapolation techniques,
such as the BST method. As known from the theory of asymptotic series
analysis, one can obtain accurate limiting value by the method if
i) the terms of the series are numerically very accurate and ii) there are
no strong confluent singularities present. To satisfy the first requirement
one needs a numerical accuracy of the data of at least $10^{-6}-10^{-7}$,
which can be achieved by the DMRG method, even at the critical point,
if moderately large finite systems are studied. These systems, as demonstrated 
by our present study, are still about one order of magnitude larger than those 
diagonalized by the L\'anczos algorithm and used in the traditional finite size 
scaling analysis.

In many physical problems the second requirement, i.e. the absence of
(strong) confluent singularities is not satisfied and these corrections to
scaling contributions represent the real limitations of the finite size
scaling method.	Since the strength of the confluent singularities decreases
with the size of the systems one expects in many cases, like to our example
with the $Q=3$ and $Q=4$ Potts models, more accurate results about the
critical exponents by the DMRG analysis compared with the traditional
finite size scaling. 

In this paper we demonstrated another advantage of the DMRG method, namely
at present it seems to be the only numerical method which could be used
to study the density profiles in critical systems, especially in the
parallel plate geometry \cite{ourPRL}. Our numerical studies on the profiles, 
together with the evaluation of the critical exponents in Sec. \ref{sec:crit}, 
have given strong numerical evidence that the conformal predictions about the
critical density profiles of the $Q$ state Potts model are exact in the
scaling limit. Our results are satisfactory, even for the $Q=4$ state
model, in which case the first order finite size logarithmic corrections
were taken consistently into account.

We mention that the analysis of this paper can be used and extended
for other problems as well. For quantum Hamiltonians one expects similar
or even better numerical accuracies, since the Hamiltonian is usually
represented by a more sparse matrix than the transfer matrix of the
equivalent classical problem. By this method one can also study such systems
in which the critical point is not known exactly by duality.
Finally one could also study first order phase transitions by the present 
approach, the results of this type of investigations will be presented 
elsewhere \cite{carlonigloi98}.

{\bf Acknowledgements } -
F.I.'s work has been supported by the Hungarian National Research Fund under
grants No OTKA TO12830, OTKA TO23642 and OTKA TO17485 and by the Ministry
of Education under grant No FKFP 0765/1997. He is indebted to the National
Committee for Technological Development for a traveling grant (COST P1) and
to the Institute for Theoretical Physics, Katholieke Universiteit Leuven, where 
part of this work has been completed, for kind hospitality. 
E.C. is financially supported by KULeuven Research Fund F/96/20.
Useful discussions with A. Drzewi\'nski, \"O. Legeza, I. Peschel, J. S\'olyom 
and L. Turban are gratefully acknowledged.

\end{multicols}
\end{document}